\documentclass{iaus}
\usepackage{graphicx}

\title[Radio Tests of GR] 
{Radio Interferometric Tests \\ of General Relativity}

\author[Fomalont \& Kopeikin]   
{Edward Fomalont$^1$ \and Sergei Kopeikin$^2$}

\affiliation{$^1$National Radio Astronomy Observatory, \\ 520
Edgemont Road, Charlottesville, VA 22903, USA \\
email: {\tt efomalon@nrao.edu} \\[\affilskip]
$^2$Dept. of Physics and Astronomy, \\ University of Missouri,
Columbia, MO 65211, USA \\email: {\tt kopeikins@missouri.edu}\\[\affilskip]}

\pubyear{2008}
\volume{xxx}  
\pagerange{119--126}
\jname{Title of your IAU Symposium}
\editors{A.C. Editor, B.D. Editor \& C.E. Editor, eds.}
\begin{document}

\maketitle

\begin{abstract}

Since VLBI techniques give microarcsecond position accuracy of
celestial objects, tests of GR using radio sources as probes of a
gravitational field have been made.  We present the results from two
recent tests using the VLBA: In 2005, the measurement of the {\it
classical} solar deflection; and in 2002, the measurement of the
retarded gravitational deflection associated with Jupiter.  The
deflection experiment measured $\gamma$ to an accuracy of $3\times
10^{-4}$; the Jupiter experiment measured the retarded term to 20\%
accuracy.  The controversy over the interpretation of the retarded
term is summarized.


\end{abstract}

\firstsection 
\section{Introduction}

The theory of general relativity (GR) describes the interaction of
matter and light with a gravitational field; hence, any accurate
measurement of this interaction is a test of GR.  The simplest test
was the GR prediction of the angular deflection of starlight passing
near the limb of the sun, first performed in 1916 during a solar
eclipse.  The results of other experiments, most using radio light
rather than star light, agreed with the GR prediction to $<0.1$\%
accuracy.  Departures of $\gamma$ from unity are expected at the
$10^{-6}$ level, and more accurate deflection observations will
continue.

Other properties of gravity can be measured with different
experiments.  For example, the perihelion shift of Mercury is a
measure of the non-linearity of the GR.  The second measurement
described in this paper, that of the retarded deflection of light
caused by the motion of the gravitating body, obtained results in
agreement with GR, but its interpretation with the property of gravity
that is constrained by this experiment is controversial.

\section {The 2005 VLBA Deflection Experiment}

\begin{figure}[t]
\vspace*{0.1 cm}
\begin{center}
 \includegraphics[width=4.1in]{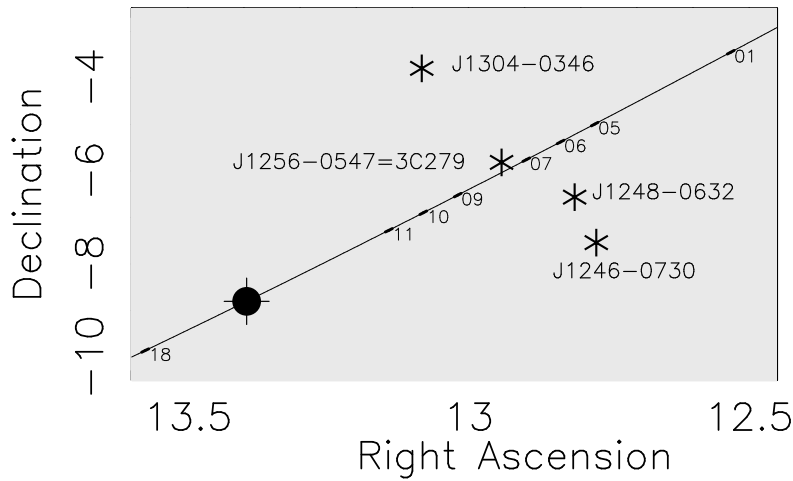} 
 \caption{Source Configuration and Observation Dates for the 2005 October Deflection Experiment: }
   \label{figfomalont1}
\end{center}
\end{figure}

We believed that a new, well-designed, VLBA experiment could
significantly improve upon the accuracy of previous solar radio
deflection experiments for two reasons.  With a stable electronic
system, sensitive receivers and accurate astrometric/tropospheric
modeling, the relative position of sources separated by a few degrees
in the sky were now being routinely measured by the VLBA with about
$20~\mu$as accuracy.  Secondly, the VLBA was now operating routinely
at 43 GHz where solar coronal effects are relatively small.  With a
typical troposphere coherence time of 1 min (maximum integration time
per observation) and the compactness of most quasars, several groups
of quasars within a five degree region near the ecliptic were easily
detectable.  The source configuration that we selected for the October
experiment is shown in Fig. \ref{figfomalont1}.

We observed on 8 days, each for a period of 5 hours.  On October 1
and 18, the relative gravitational and coronal bending was less than 2
$\mu$as so the undeflected relative positions for the sources could be
determined.  Also, any significant structural changes of the quasars
over the experimental period could be ascertained (none were seen).
Significant relative gravitational bending was obtained on the other
observing days.  The observations on each day were identical.  Each
source was observed for 45 sec with a switching time of about 25 sec,
in the sequence: 3C279, J1304, 3C279, J1256, 3C279, J1258, 3C279,
J1304, etc.  As long as the phase between subsequent 3C279
observations could be connected, this observing scheme accurately
determined the source relative positions.

In order to determine the coronal refraction at 43 GHz, we time-shared
the observations on each day among 43, 23, and 15 GHz.  This was done
by using the source observing sequence noted above, but switching
frequencies every 25 minutes.  In this manner we determined the
relative position of the sources at any one frequency in each 25-min
block to an accuracy of $\sim 100~\mu$sec.  By comparing the source
positions as a function of frequency, we determined any significant
frequency dependent position change.  However, when a source was
closer than about $2.2^\circ$ of the sun, the coronal refraction {\it
vibrated} the source position at 43 GHz more than $50~\mu$sec over a
time scale of a few seconds, making the phase tracking impossible to
follow.  Hence, all data were phase unstable for October 7 and several
sources could not be used on October 5, 6 and 9.  An additional 10\%
of the data were lost for some VLBA telescopes during periods of poor
weather.

As a rough guide to the experiment sensitivity, at 43 GHz the typical
accuracy for the relative position between two sources over one five
hour period was about 0.07 mas.  With a relative gravitational
deflection 100 mas for most observing days, we obtained a deflection
accuracy of 1 part in 1400 (0.0007) for each source-pair and day.
When all source-pairs and are averaged at 43 GHz, we obtain
$(\gamma-1) = -0.00070\pm 0.00040$ (rms error).  If we include the 23
GHz and 15 GHz data to remove the small coronal bending, we obtain
$(\gamma-1) = -0.00006\pm 0.00027$, our best estimate.

We believe that we can increase the experiment accuracy by a factor
two to three.  First, a group of sources in May when the sun is at a
more northern declination will improve the astrometric quality of the
data.  Second, the most accurate results are obtained when the sources
are between $3^\circ$ to $5^\circ$ from the sun.  Third, the 2005
experiment scheduled too much time with sources too near the sun and
not enough time at these intermediate solar separations.  Finally,
most of the observation time should be made at 43 GHz, with perhaps
25\% of the time at 23 GHz to remove the significant coronal bending.

\section {The 2002 Jupiter Deflection Experiment}

\begin{figure}[t]
\vspace*{0.3 cm}
\begin{center}
 \includegraphics[width=3.2in,angle=000]{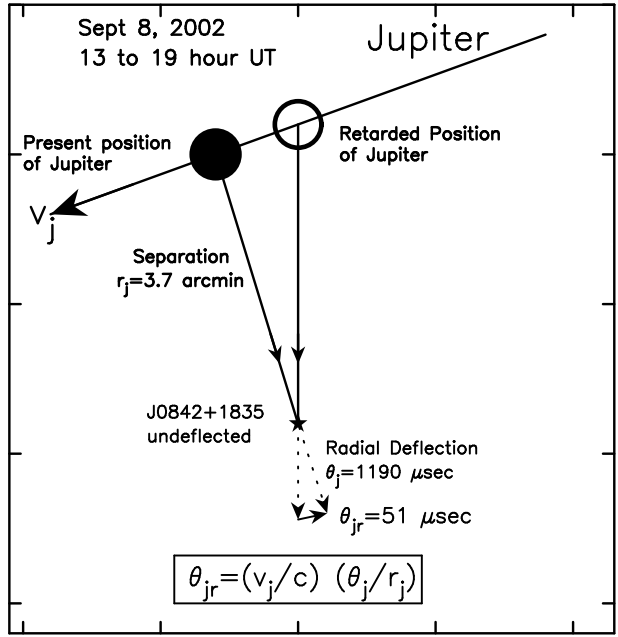} 
\vspace*{0.3 cm}
 \caption{The Retarded Deflection of the Jovian Experiment in 2002: The line shows
the path of Jupiter during the Sept 8 observations, with closest approach at
16:30 hr UT.  The relevant deflection angles are shown.}
   \label{figfomalont2}
\end{center}
\end{figure}

On September 8, 2002, Jupiter passed within $3.7'$ of the quasar
J0842+1835.  Such a close passage of Jupiter with a bright quasar (0.4
Jy, among the brightest 1000 sources) occurs on average once every 20
years.  A similar encounter in 1988 with a different bright quasar was
observed, and the gravitational deflection of about 1 mas was detected
by \cite[Treuhaft \& Lobe (1991)]{tru91}.  Our goal was to measure not
only the radial deflection, but the retarded component as well.  The
relevant parameters at closest approach are shown in
Fig. \ref{figfomalont2}.  The maximum radial deflection was
$1190~\mu$as and occurred at 16:30 UT on September 8.  Because of the
motion of Jupiter, there is a retarded deflection component of
$51~\mu$as opposite to the direction of motion of Jupiter, and is
within reach with the VLBA.

The details of the experiment and the results have been published
by \cite[Fomalont \& Kopeikin (2003)]{fom03}.  The VLBA observed
at 8.4 GHz and observations switched between J0842+1835 with
another quasar about $0.8^\circ$ east J0839+1802, and with J0854+2006,
about $3.4^\circ$ to the west.  A complete cycle took about 5 minutes.
Five observing days, each 7 hours long, were made on September 4, 7,
8, 9, 12.  Since the retarded term was significant only on September 8, the
other four days were used to measure the undeflected position of the
sources and to determine realistic errors.   

These three quasars were positioned nearly linearly in the sky; hence,
the appropriate combination of the measured phases for J0839 and J0854
not only removed the temporal troposphere and ionosphere refraction
changes, but also the effect quasi-stationary phase gradients at the
position of J0842.  These systematic phase gradients are caused by
many small astrometric effects (antenna location offsets,
earth-orientation modeling errors), as well as troposphere and
ionospheric structure from two to twenty degrees in the sky.  The use
of two calibrators rather than one reduced the residual position
uncertainty per day from about $25~\mu$as to about $10~\mu$as
(\cite[Fomalont (2003)]{fom03a}).

Both the radial and retarded deflections were easily detected.
Analysis of the radial deflection gave $\gamma=1.01\pm 0.03$.  The GR
prediction of the retarded deflection varied between 41 and $51~\mu$as
on September 8, and the ratio of the measured retarded deflection to
the GR prediction (assumes that the velocity associated with the
retardation is the speed of light) was $0.98\pm 0.19$.

The experiment confirms the GR prediction for the retarded deflection
at the 20\% level.  A useful question is: what property of gravity is
constrained by this experiment?  Our interpretation is that the
retardation is a measure of the propagation speed of gravity, and is
related to a gravito-magnetic field associated with currents (motion)
of matter (\cite[Kopeikin \& Fomalont (2007)]{kop07}).

A summary of other interpretations has been compiled by \cite[Will
(2008)]{Wil08}.  The basic disagreement is whether the speed of
propagation of light or gravity (c) can be manifested in (v$_j$/c)
terms or only in (v$_j$/c)$^2$ terms, where v$_j$ is the velocity
relevant to the object.  More specifically, \cite[Asada (2002)]{asa02}
claims that the speed of light was determined. \cite [Will
(2003)]{wil03} believes that the PPN parameter $\alpha_1$ was measured,
albeit poorly. \cite[Carlip (2004)]{car04} finds the distinction
between the speed of gravity and the speed of light somewhat
ill-posed.  \cite [Stuart (2004)]{stu04} believes that the experiment
measured nothing useful.

\section {Summary}

The two experiments demonstrate that the micro-arcsecond positional
accuracy of the VLBA produces significant tests of GR.  The goal of
the tests were to measure important parameters and their limits, and
to foster discussion concerning the generalization of GR and its
interaction with light and mass.

The National Radio Astronomy Observatory is a facility of the National
Science Foundation operated under cooperative agreement by Associated
Universities, Inc.


\begin{thebibliography}{}

\bibitem[Asada (2002)]{asa02}
{Asada, H. 2002}, \textit{ApJ}, 574, L69

\bibitem[Carlip (2004)]{car04}
{Carlip, S. 2004}, \textit{Class. Quant. Grav.}, 21, 3803
(arXiv:gr-qc/0403060v3)

\bibitem[Fomalont (2003)]{fom03a}
{Fomalont, E.~B., 2003}, \textit{Future Directions in High Resolution Astronomy}, Ed. J. D.Romney
and M. J. Reid, Socorro, NM, p55

\bibitem[Fomalont \& Kopeikin (2003)]{fom03}
{Fomalont, E.~B. \& Kopeikin, S.~M. 2003}, \textit{ApJ}, 598, 704

\bibitem[Kopeikin \& Fomalont (2007))]{kop07}
{Kopeikin, S.~M. \& Fomalont, E.~B. 2007}, \textit{Gen. Rel. and Grav.}, 39, 1583
(arXiv:gr-qcd/0510077v4)

\bibitem[Stuart (2004)]{stu04}
{Stuart, S.  2004}, \textit{Int. J. Mod. Phys.}, D13, 1753.

\bibitem[Treuhaft \& Rowe (1991)]{tre91}
{Treuhaft, R.~N., \& Lowe, S.~T. 1991}, \textit{AJ}, 102, 1879

\bibitem[Will 2003]{wil03}
{Will, C.~M., 2003}, \textit{ApJ}, 590, 683

\bibitem[Will 2008]{wil08}
{Will, C.~M., 2008, http://wugrav.wustl.edu/people/CMW/SpeedofGravity.html} 

\end{thebibliography}
\end{document}